# Virtual images and billiards

J. Christopher Moore, Richard D. Floyd and Cody V. Thompson

*Department of Chemistry and Physics*

*Coastal Carolina University*

*P.O. Box 261954, Conway, SC 29528*

*moorejc@coastal.edu*

Students in introductory physics courses struggle to understand virtual image formation by a plane mirror and the proper construction of ray diagrams.[1] This difficulty, if not sufficiently addressed, results in further problems throughout the study of geometric optics. Specifically, students fail to apply proper graphical representation of light rays during investigations of the formation of real images by converging lenses and concave mirrors.[2] We present a guided-inquiry activity based on the research-verified *Physics by Inquiry* text[3] that incorporates a small and inexpensive billiards table, with billiard balls acting as "light". In this way, we approach the abstract concept of virtual images by relation to the concrete concept of physical reflection.

Virtual images form behind the mirror plane at the intersection of hypothetically extended reflected rays. David Singer demonstrates this using a lit candlestick,[4] whereas Peter Mansell employs the student's finger poked through the hole of a compact disc.[5] Thomas Greenslade addresses the formation of the image of a virtual image in a plane mirror by use of a "Hall of Mirrors."[6] Using traditional techniques, we have found post-instruction that students can correctly identify the location of a virtual image with little difficulty when the mirror is sufficiently large in comparison to the object. However, it appears this knowledge is obtained by rote and requires mostly descriptive level reasoning, since when more challenging questions are posed involving fixed-size mirrors with extended objects, students still struggle. This observation is consistent with previous pedagogical research on student misconceptions in geometric optics.[1]

Student difficulties with the concept of virtual images may result from poorly developed high-order scientific reasoning. Specifically, in conceptual physics courses for non-scientists, abstract concepts such as force, energy and virtual images pose a significant challenge to students due to inadequate preparation in scientific reasoning and a reliance on mostly descriptive thinking.[7,8] We attempt to ease students into the abstract concept of virtual images via the more concrete example of physical reflection of billiard balls.

Figure 1 shows (a) a side view and (b) a top view of a miniature billiards table containing two billiard balls and a plane mirror. Students are instructed to look at the black 8-ball through the mirror [similar to fig. 1(a)] and determine the image's location via the method of parallax.[9] They then proceed to determine the location of the other features of the billiards table, such as the side and corner pockets and the white cue ball. They draw this image on a large piece of white paper immediately next to the billiards table. The resulting image resembles that shown on the right in fig. 1(b). Before this exercise, students will typically be familiar with the notion that images are on the opposite side of the mirror and the same distance away as the object. However, this initial billiards table activity helps combat the misconception that objects not immediately in front of the mirror have no image, which is a commonly held belief with fixed-size mirrors even after traditional instruction.

Students then use a cue stick to strike the cue ball, attempting to "bank" the cue ball off of the sidewall and knock the 8-ball into the side pocket. Some students instinctively aim for the 8-ball that they see in the mirror; however, all students eventually discover via trial and error that aiming for the image results in success. Students are instructed to carefully observe the path of the ball both incident and reflected and mark this path on the table using either a colored tape or chalk. They then use the method of parallax to determine the image location for several points along these lines. The result is the image shown in fig. 1(b). Protractors are used to show that the angle of incidence is equal to the angle of reflection.

Next, we have students repeat the procedure outlined above, however, we have them move the cue ball to a different location. This is analogous to observing an image from two different locations, which is necessary to construct a useful ray diagram that illustrates the image location. The two straight-line motions through the mirror intersect at the image, guiding students to an operational procedure for determining image location. After this exercise, we typically continue through the activities as written in *Physics by Inquiry* until we reach the sections dealing with multiple reflection, at which point we return to the billiards table.

Figure 2 shows (a) a side view and (b) a top view of the same miniature billiards table containing two billiard balls and two plane mirrors. Students proceed as before; however, in this case, they must bank the cue ball off of both sidewalls in order to successfully knock the 8-ball into the corner pocket. In this case, students begin the examination of an image of a virtual image, which is a terrifically difficult concept for students to grasp. They construct two images fields [see fig. 2(b)], one for each mirror, and demonstrate to themselves that the same principles apply for images of virtual images.

We began implementing this activity to add an exciting and enjoyable element to geometric optics, since students grow bored of traditional pins and cardboard. Plus, teaching students how to be better billiards players helps with recruitment. However, we have found via post-instruction student interviews and on open-ended examination questions that students appear to be grasping the virtual image concept more deeply. Furthermore, they struggle less during the completion of the activities in *Physics by Inquiry*. We attribute this to the activity's concretization of an abstract concept, which helps ease students transition from descriptive-level reasoning into the hypothetical realm.[8]

Figure 1: (a) Side view and (b) top view of a miniature billiards table containing two billiard balls and a plane mirror.

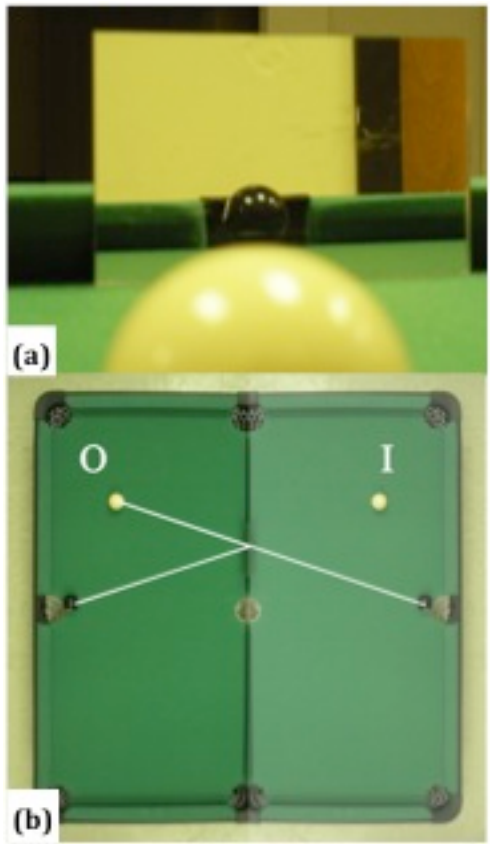

Figure 2: (a) Side view and (b) top view of a miniature billiards table containing two billiard balls and two plane mirrors.

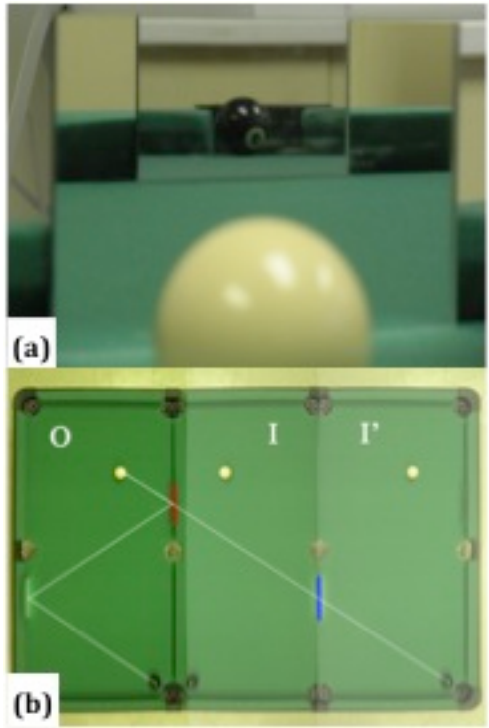